# Abstractions, algorithms and data structures for structural bioinformatics in *PyCogent*

Marcin Cieślik,[a] Zygmunt S. Derewenda[b] and Cameron Mura [a]*

[a]*Department of Chemistry; University of Virginia; Charlottesville, VA 22904, and*

[b]*Department of Molecular Physiology & Biological Physics; University of Virginia Health Sciences Center; Charlottesville, VA 22908, USA.*

*E-mail: cmura@virginia.edu*

## Abstract

To facilitate flexible and efficient structural bioinformatics analyses, we have introduced new functionality for 3D structure processing and analysis into *PyCogent* – A popular, feature-rich framework for sequence-based bioinformatics, but one which has lacked equally powerful tools for handling stuctural/coordinate-based data. We have developed extensible Python modules that provide object-oriented abstractions (based on a hierarchical representation of macromolecules), efficient data structures (*e.g.*, kD-trees), fast implementations of common algorithms (*e.g.*, surface area calculations), read/write support for PDB-related file formats, and wrappers for external command-line applications (*e.g.*, *Stride*). Integration of this code into *PyCogent* is symbiotic, allowing sequence-based work to benefit from structure-derived data and, reciprocally, enabling structural studies to leverage *PyCogent*'s versatile tools for phylogenetic and evolutionary analyses.

## 1. Introduction & Motivation

Structural biology continues to enjoy unprecedented rates of data accumulation. Together with structural genomics, advances in macromolecular crystallography have contributed to $>7\times10^4$ known 3D structures, at recent rates of several thousand new structures per year. (Terwilliger *et al.*, 2009) This deluge of coordinate data impacts fields ranging from molecular biophysics (*e.g.*, elucidating relationships between conformational dynamics $\leftrightarrow$ protein function) to structural bioinformatics (*e.g.*, determining fractional coverage of fold space) to more applied fields such as drug design (*e.g.*, incorporating new structural knowledge in designing receptor•ligand complexes). Much of the effort in these areas now involves grappling with the sheer volume and heterogeneity of data. For instance, a biochemist seeking to analyse a 'representative' set of structural data may need to embark upon the tedious task of constructing a non-redundant subset of coordinates by, perhaps, removing structures exceeding some sequence (or structural) similarity threshold; similarly, in structural bioinformatics, sufficient statistical sampling entails calculations over as large a dataset as possible, therefore necessitating robust, automatable data-processing.

Robust processing of coordinate data places stringent demands on both the data and the data-processing tools: datasets must be structured in some organized, well-defined (parse-able) manner, and the applied tools must be *scalable* (to cope with data volume), *robust* (to handle data heterogeneity, formatting errors), and *flexible* (to permit custom analyses). These requirements arise regardless of the sophistication or computational complexity of the structure analysis being undertaken.

### *1.1. Data Format Difficulties*

In terms of data format, the Protein Data Bank (PDB) file (Bernstein *et al.*, 1977) is the standard exchange and archival system for macromolecular structures (atoms and 3D coordinates) and associated structure-determination data (biopolymer sequence, details of the diffraction or NMR experiment, refinement steps, *etc.*). More flexible data exchange formats, such as mmCIF (Westbrook & Bourne, 2000), have been devised, as have application-specific derivatives of the PDB format – *e.g.*, 'PQR' files (Dolinsky *et al.*, 2004) replace the $B$-factor and occupancy fields with atomic charges ('$Q$') and radii ('$R$'), for use in electrostatics calculations . However, structure analysis schemes based on replacing standard PDB fields are inherently limited: *(i)* custom scripts are generally required to apply user-defined(/computed) values on a per-atom basis (*i.e.*, mapping onto '`ATOM`' fields), rather than at an arbitrary level in the structural hierarchy (per-residue, per-chain, *etc.*); and *(ii)* this approach is limited to storage of only a couple different types of (scalar) data fields (*e.g.*, $Q$, $R$). Nevertheless, the PDB file format remains the *de facto* standard for biopolymer 3D structures, and therefore must be dealt with both flexibly and robustly in order to automate structural analysis workflows.

By providing high-level abstractions and robust data structures for coordinate (and coordinate-derived) data, the software introduced here transcends the difficulties and limitations of such approaches as manipulating fixed-width, predefined PDB fields.

### *1.2. Data Processing Challenges*

Typical structure analysis pipelines begin with 3D coordinates as primary data, and involve computing a wide variety of secondary/derived properties. Calculated quantities may include *(i)* physics-based characteristics, such as electrostatic potentials; *(ii)* geometric features, such as concavity and curvature, shape complementarity indices (Lawrence & Colman, 1993), surface areas (buried, solvent accessible, Van der Waals envelope, *etc.*); *(iii)* protein structural descriptors (2° structure assignments); and *(iv)* a host of 'hybrid' physicochemical measures, such as hydrogen-bonding patterns, geometric distributions of residues (*e.g.*, radial distribution functions of atom types), hydrophobic clusters, interaction patches, hydrophobic moments, and so on. Many separate tools for performing one or the other of these sorts of calculations exist, but often are in the form of machine/architecture-specific binaries that take some input files and command-line arguments, and yield resulting data as output files in a format highly specific to that particular program. [Indeed, one of the early motivations for software suites such as CCP4 was to unify the input/output syntax across a set of individually-useful programs (Collaborative Computational Project, 1994).] However, performing structure analysis in this manner is brittle, error-prone, and almost always necessitates development of a set of post-processing scripts ('glue-code') to finesse the results into a form usable for the next stage of data analysis. This approach becomes further untenable if multiple streams of output must be combined (extensive dataflow interdependencies), or if the structure analysis steps require chaining together numerous output values $N_{i-1}{\rightarrow}N_i{\rightarrow}N_{i+1}$. In short, 'one-off' tools/scripts are generally quite idiosyncratic in nature, with each new set of analyses requiring a new set of scripts.

The software presented here provides general-purpose libraries and tools – easily accessible and manipulable data structures, algorithms for coordinate-based calculations – for both routine and sophisticated (user-configurable) structural analyses. Integration of this functionality into the versatile, well-established *PyCogent* bioinformatics framework (Knight *et al.*, 2007) effectively supplies a highly extensible toolkit for structural bioinformatics workflows, obviating many of the data-processing challenges described above.

## 2. Overview of the Developed Software

After outlining the architecture, design considerations, and implementation of our code, this section summarizes general software capabilities. Thorough descriptions of our abstractions and data structures (illustrated with concrete examples), as well as notes on highly specific details such as our code's handling of sequence and structural heterogeneity in PDB files (atomic alternate location identifiers, insertion codes), can be found in the *Supplementary Material* (S§1-2).

### *2.1. Architecture & Design*

We have employed a hierarchical internal representation of macromolecular structures, reflecting the implicit *Structure* $\supset$ *Model* $\supset$ *Chain* $\supset$ *Residue* $\supset$ *Atom* ('*SMCRA*') organization of a PDB file (Hamelryck & Manderick, 2003). Such a hierarchy is most naturally programmed in the object-oriented paradigm. Elements of our hierarchical design pattern are termed '*entities*', with base classes '*Entity*' and '*MultiEntity*'. At the top-tier of the hierarchy is a '*Structure*', which is a container for '*Models*'. Each *Model* can hold multiple '*Chains*', which in turn contains '*Residues*'. The bottom of the hierarchy consists of '*Atoms*', which are themselves individual entities (not higher-order; childless). Thus, any atom stored in our data structure is uniquely identified by its location within the hierarchy, effectively providing a per-atom *composite identifier*. Each entity between the (top) level of *Structure* and the (bottom) level of *Atom* is linked to potentially multiple children (entities lower in the hierarchy tree) and to only a single parent (higher in the hierarchy tree). In other words, there are no cycles or closed loops in the graph of this tree-like hierarchy. As a concrete example, note that a particular *Residue* has a single *Chain* parent, and multiple *Atom*ic children.

We extend this *SMCRA* schema in *PyCogent via* an additional group of virtual "holder entities" lying outside the hierarchy. These *holders* are used to define groups of entities irrespective of their location within the hierarchy, thereby enabling one to combine structural units (residues, atoms) into application-specific clusters, patches, or any other type of arbitrary (user-definable) collection of structural elements. Each *holder* is technically an unrestricted grouping of entities of the same hierarchical level (*e.g.* atoms), and holder children are guaranteed to have at most one parent. Holders, which may be (loosely) thought of as "atom selections" in the macro language of programs such as *PyMOL* (DeLano, 2002), are intended to be temporary in nature, and can be created or destroyed without affecting the underlying macromolecular structure. In the parlance of object-oriented programming, the *Entity* is a *base class* (or a '*superclass*'), and the relationship between the two complementary hierarchies – *SMCRA* and holders – is most definitively seen by inspecting the *Entity* inheritance diagram.

*PyCogent* abstracts a PDB file as a top-level *Structure* entity in the hierarchical data structure described above, and allows for its easy and arbitrary manipulation without compromising the uniqueness of the corresponding structure. As explicitly shown in some of the examples in the supplementary material (S§2), modifications to the *Structure* can involve such operations as changing the attributes drawn from the PDB file (*e.g.*, altering coordinates or residue numbers), or annotating with additional attributes that are generated either within *PyCogent* proper or by external libraries (*e.g.*, computing electrostatic potentials and mapping the values as attributes of the nearest atom to a voxel). Such modifications are typically on per-atom or per-residue bases, and are therefore initially mapped onto entities at those levels. However, *PyCogent* also provides functionality to propagate these values across the hierarchy tree. The 'type' of propagation that is performed depends on the nature of the data items – *(i) scalar* values can be summed or averaged (*e.g.*, the accessible surface area (ASA) of a residue is the sum of ASA values of its constituent atoms, whereas residue depth could be computed as mean atomic depth), while *(ii) ordinal* data types are generally propagated using set-theoretic (union or intersection) operations (*e.g.*, defining the list of atoms contacting residue $i$ as the union of all the pairwise contacts of atoms in residue $i$ with non-$i$ atoms).

### 2.2. Implementation & Availability

Implemented in standard (CPython), platform-independent Python, our software makes extensive use of object-oriented (OO) programming. The benefits of scientific software development in the high-level, interpreted Python language, using an OO approach, have been reviewed elsewhere (Knight *et al.*, 2007) (Grosse-Kunstleve *et al.*, 2002). Key OO concepts in our codebase – the *Entity* base class, attribute dictionaries, 'holders', *etc.* – are outlined above and further described in S§1-2 of the *Supplementary Materials*. In addition, section S§3 of the *Supp Materials* provides further information on code performance (benchmarking and timing statistics [S§3.1]) as well as notes on platform independence and Cython extensions ([S§3.2]). The code development described here has been integrated into the latest stable release (v1.4.1) of the open-source *PyCogent* project, and is freely available at http://pycogent.sourceforge.net. Extensive documentation can be found at that URL and in the *Supp Materials*.

### 2.3. Capabilities & Features

The software introduced herein extends *PyCogent*'s capabilities into the realm of crystal structure processing and analysis (Table 1). The application programming interface (API) aims to strike a balance between flexibility/extensibility (advanced users, more complex analyses), *versus* robust/simpler default behavior for handling of PDB files and PDB-derived structural data. Thus, various capabilities are provided by generalized classes and methods, which expose that particular functionality (*e.g.*, surface area calculations) to advanced users with as few limitations as possible. For example, the default parameter set of atomic radii that we provide, drawn from the *Areaimol* program (Collaborative Computational Project, 1994), can be easily user-adjusted for different applications and purposes.

In addition to reading/writing standards-compliant PDB files (and close relatives, such as PQR), we provide facilities for (1) arbitrary entity selections (atoms, residues, *etc.*); (2) flexible data manipulation and propagation (described above); (3) protein structure clean-up; (4) fast calculation of accessible surface areas (ASA) and molecular surfaces; (5) unit cell- and lattice-related calculations (finding crystal contacts, computing coordination numbers); (6) nearest neighbor searches; and (7) superimposing structures. Some of these functionalities are more computationally expensive, be it due to numerical reasons (*e.g.*, surface area calculations) or combinatorial complexity (*e.g.*, constructing nearest-neighbor and contact lists); computationally intensive portions of the code were implemented using C extensions for Python, *via* the Pyrex-based Cython compiler (Behnel *et al.*, 2008). Further details, particularly with respect to issues of cross-platform compatibility and C extensions, are provided in S§3 of the *Supp Materials*.

Most of the aforementioned features are intrinsic (*i.e.*, free of external dependencies), including surface area calculations and contact searching (Table 1). Parameters for each intrinsic algorithm are user-accessible and adjustable – one can modify atomic and probe radii, source and target atom collections for contacts, *etc.* The algorithms properly handle crystal lattices and space group symmetry such that, for instance, the correct neighborhood of symmetry-mates are generated in crystal contact calculations. Other functionality is provided by wrapping existing, freely-available software – for instance, application controllers and utility functions are provided for molecular surface calculations using *MSMS* (Sanner *et al.*, 1996) and 2° structure assignment *via* *Stride* (Frishman & Argos, 1995). In such cases, all input files required by the external binary are auto-generated from our hierarchical representation, and users can modify command-line switches and parameters; finally, we provide parsers for the resulting output, that can then be used to annotate the structure under study. The following subsections describe implementation details for the numerically-intensive functionalities – neighbor search, surface calculations, and contact calculations.

*2.3.1. Nearest neighbor searches.* Nearest neighbor search (NNS) algorithms are used in many fields, ranging from data compression to DNA sequencing and genome assembly. The $k$-nearest neighbors problem is to locate the $k$ closest points to a query point in some well-defined (*e.g.*, Euclidean, Manhattan) metric space. By employing specialized data structures and space-partitioning methods (such as kD trees), the computational complexity of NNS can be reduced from the brute-force $\mathcal{O}(N^2)$ to $\mathcal{O}(N \log N)$. Our code supplies built-in $k$-NN search procedures for arbitrary structural entities (atoms, residues, user-defined groups of residues, *etc.*) *via* a custom kD-tree implementation; this provides the basis for our software's higher-level functionality, such as calculation of entity coordination numbers, surface patches, *etc.* The kD tree functionality was written as compiled Cython code (the 'ckd3' module in the codebase).

Major functionality and specific utilities provided by our code are indicated as 'intrinsic'; tools provided *via* external packages are referenced in the text.

| | |
|---|---|
| Read, write, and manipulate PDB files (standardize, renumber, modify fields, *etc.*) | intrinsic |
| Miscellaneous geometric analyses (*e.g.*, compute geometric centers) | intrinsic |
| Nearest neighbor search, $k$D trees | intrinsic |
| Interatomic contact calculations | intrinsic |
| Molecular surfaces | *MSMS* |
| Accessible surface area (ASA) calculations | intrinsic, *Stride* |
| Atom/residue depth calculations (under development) | intrinsic |
| 2° structure calculation and residue-based ASA | *Stride* |
| Proper handling of space group symmetry | intrinsic |

*2.3.2. Surface calculations.* Of the many physicochemical features of a protein structure, the accessible surface area (ASA) plays a key role in defining biochemical activity. The new code reported here now provides the *PyCogent* package with many tools (again, intrinsic and external) for computing and handling surface area data. The relative accessible surface areas (rASA) can also be computed; this paramter is normalized by the maximum possible ASA (*i.e.*, in an extended peptide) for each amino acid, thereby offering a useful criterion for comparison between different residue types (Rost & Sander, 1994). Residue-based ASAs can be computed using our interface to *Stride*; though primarily included for purposes of 2° structure assignment, *Stride*-derived ASA values can be used for cross-validation and comparison, *versus* ASA's computed using our built-in (Cython-encoded) ASA facility. (Thus, users are provided with multiple options for a single type of task.) An interface is also provided to the *MSMS* program, which exists as a standalone binary that takes as input a file of coordinates and associated atomic radii, and generates an output file in the form of an array of surface dots. Given a PDB file our software *(i)* produces the necessary input files for an *MSMS* calculation, *(ii)* allows user modification of such parameters as atomic radii and run-time options (surface accuracy), and, finally, *(iii)* provides data structures to access the output array as a matrix of floating point numbers.

*2.3.3. Intermolecular contacts.* Several types of intermolecular contacts may be important in biochemical and structural studies of physiological functions (protein···protein, protein···ligand interactions); in characterization of biophysical properties, such as hydration (protein···water); and in understanding the crystallization process itself (homologous and possibly heterologous protein···protein contacts within an asymmetric unit and across a lattice). The problem of computing chain or residue coordination numbers is an essentially identical task as contact analysis. The coordinates in a PDB file correspond to a single asymmetric unit, posing a potential problem for crystal contact calculations. Our *PyCogent* implementation provides functionality for inter-*entity* (inter-{atom, residue, chain}) calculations, as detailed in S§4 of the *Supp Materials*; notably, lattices and space group symmetry are properly handled such that, for instance, the correct neighborhood of symmetry mates is generated in *PyCogent*-based crystal contact calculations.

## 3. Case Studies

The first case study below (§3.1, Fig 1) shows the ease with which routine structural tasks can now be performed in *PyCogent*. Then, to demonstrate how our new software can leverage *PyCogent*'s existing repertoire of sequence-based functionality to perform more complex structural analyses, an advanced example (§3.2, Fig 2) computes sequence conservation scores (as profile entropies) and maps the resulting information onto a structure. Additional examples are provided in the *Supp Materials*, including *(i)* further details on this second use-case (S§5.1); *(ii)* calculation of interatomic contacts (S§4); and *(iii)* a use-case that expands upon the §3.2 sample shown below by demonstrating how coordinate-derived data (*e.g.*, 2° structural information) can be incorporated into a relatively sophisticated statistical sequence/structure calculation. Notably, such a calculation cannot be performed as efficiently(/compactly) in existing structural analysis software packages of which we are aware.

### *3.1. Display the Quaternary Structure of an Oligomer*

To demonstrate the ease of accessing our new structural tools in *PyCogent*, the following block of code shows, in conjunction with a Python-aware molecular graphics system, how one can efficiently (∼10 lines) go from a raw PDB file to a visual representation of the quaternary architecture of a protein complex. In this example, the center of geometry is automatically computed (as implicit attributes of *Chain*-level entities) by *PyCogent*'s structure-handling functionality. PyMOL's compiled graphics object (CGO) facility is utilized to render the centers as spheres (Fig. 1), the final CGO object having been built-up *via* successive iterations over the chains of the assembly (line 4 loops at level '*C*' of *SM**C**RA*). Note that the following block of code is entered directly at the PyMOL command-prompt (a further note on setting-up PyMOL to be *PyCogent*-aware can be found in S§3.3 of the *Supp Materials*):

```
from cogent.parse.pdb import PDBParser # PyCogent
myPDBfile = open('./1i8f.pdb')
myNewStruc = PDBParser(myPDBfile)
for chain in myNewStruc.table['C']:
  myChainID = myNewStruc.table['C'][chain].getName()
  # initialize a PyMOL CGO list:
  cgoString = [COLOR, 0.7, 0.7, 0.7]
  if myChainID != ' ':    # only non-empty chains
    # fetch center of geometry as numpy array:
    center = myNewStruc.table['C'][chain].coords
    # build-up CGO:
    cgoString.extend([SPHERE, center[0], center[1],\
        center[2], 3.0])  # 3 A-radius sphere
    # and load into PyMOL as uniquely id'd spheres:
    cmd.load_cgo(cgoString,'cent_'+myChainID)
```

Fig 1: Architecture of a heptameric assembly (PDB 1I8F), illustrated in terms of the geometric centers of the entire complex (grey sphere) and its individual subunits (tan spheres). Automatically-computed center positions were retrieved while looping over protein chains (see code in text).

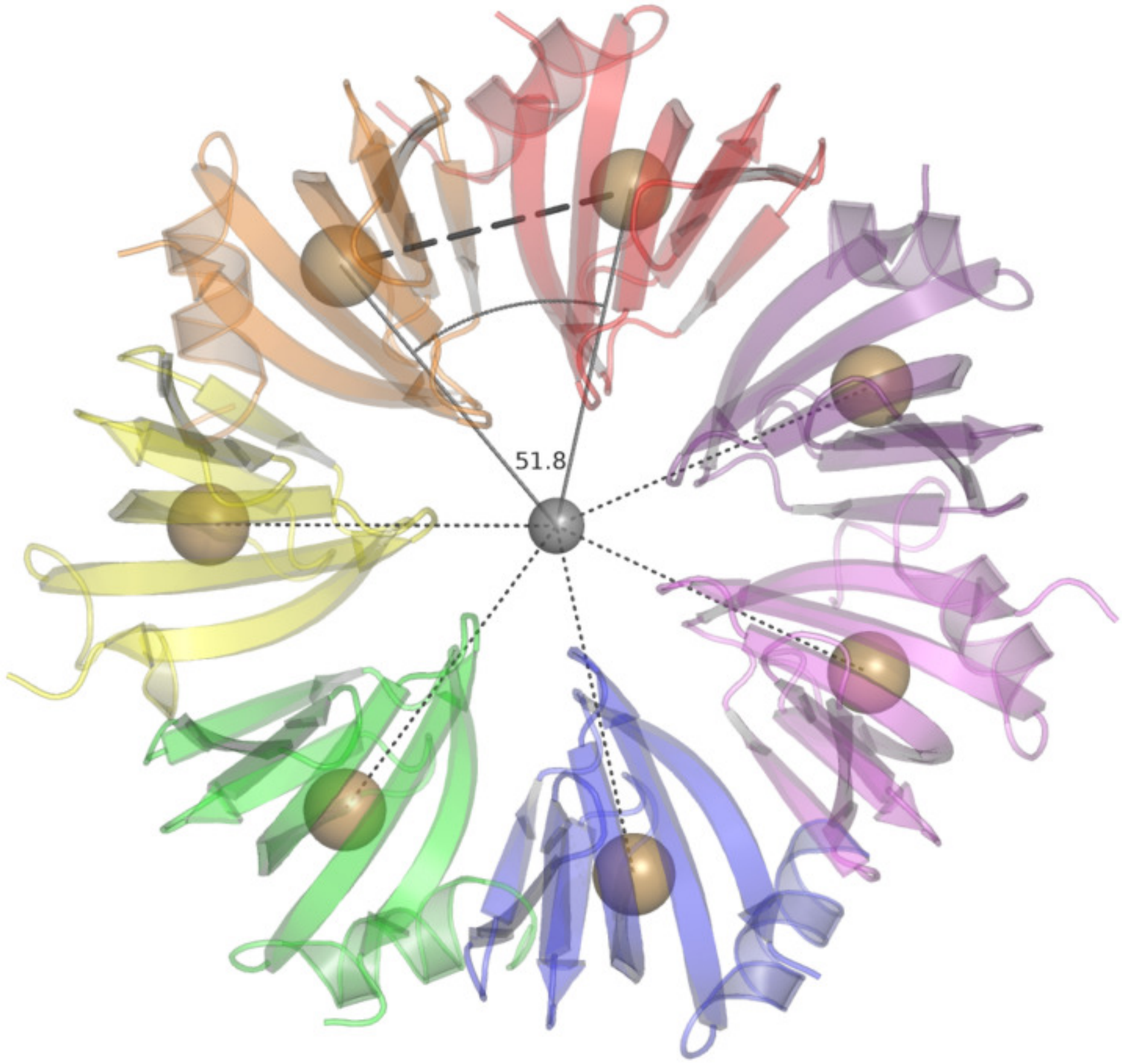

*3.2. Compute Sequence Entropy, Map onto Structure*

This case study illustrates how sequence-derived information can be mapped onto a structure, *e.g.*, for purposes of visualization. The specific task is to identify, and then illustrate, fragments of a protein structure that are conserved at the level of sequence, with 'conservation' measured as bits of entropy (a quantity readily computable from an alignment; see, *e.g.*, (Durbin *et al.*, 1998)). This analysis proceeds through several stages, involving *(i)* loading an alignment and computing its profile-based Shannon entropy at each residue position; *(ii)* loading a 3D structure representative of the sequences in the above alignment, fetching (*e.g.*, from UniProtKB) sequences similar to that in the PDB file, and performing the many manipulations required to properly align it; *(iii)* compute a valid mapping of entropy scores to residues in the PDB file, in order to be able to *(iv)* annotate the 3D structure with entropy scores, and store the structure for downstream visualization (*e.g.*, by color-ramping residues based on bit scores).

Perhaps the most difficult step in the above workflow is correctly mapping residues from the 3D structure to positions in the sequence alignment. As detailed in the *Supp Materials* (S§5.1), global sequence alignment *via* the Needleman-Wunsch algorithm was used to establish a correspondence between 3D structure (PDB-derived sequence) and sequence profile position. Uncertainty bits (entropy scores) for alignment positions that correspond to gaps in the 3D structure-derived sequence (*i.e.*, missing residues in the PDB file) were simply discarded, as they cannot be visualized. The final sequence entropy $\rightarrow$ structure mapping is shown in Figure 2. Notably, integrating our structure-handling software with *PyCogent*'s bioinformatics utilities enabled this relatively involved structural analysis to be achieved fairly efficiently (<50 lines of code; see S§5.1).

Fig 2: In this example (§3.2), sequence entropy was computed for the chorismate mutase family and mapped to structure 1UI9. Note that higher-entropy segments (red/thicker tubes) correspond to loops or exposed regions, whereas the $\beta$-sheet core is characterized by lower overall entropy values (blue/thin tubes).

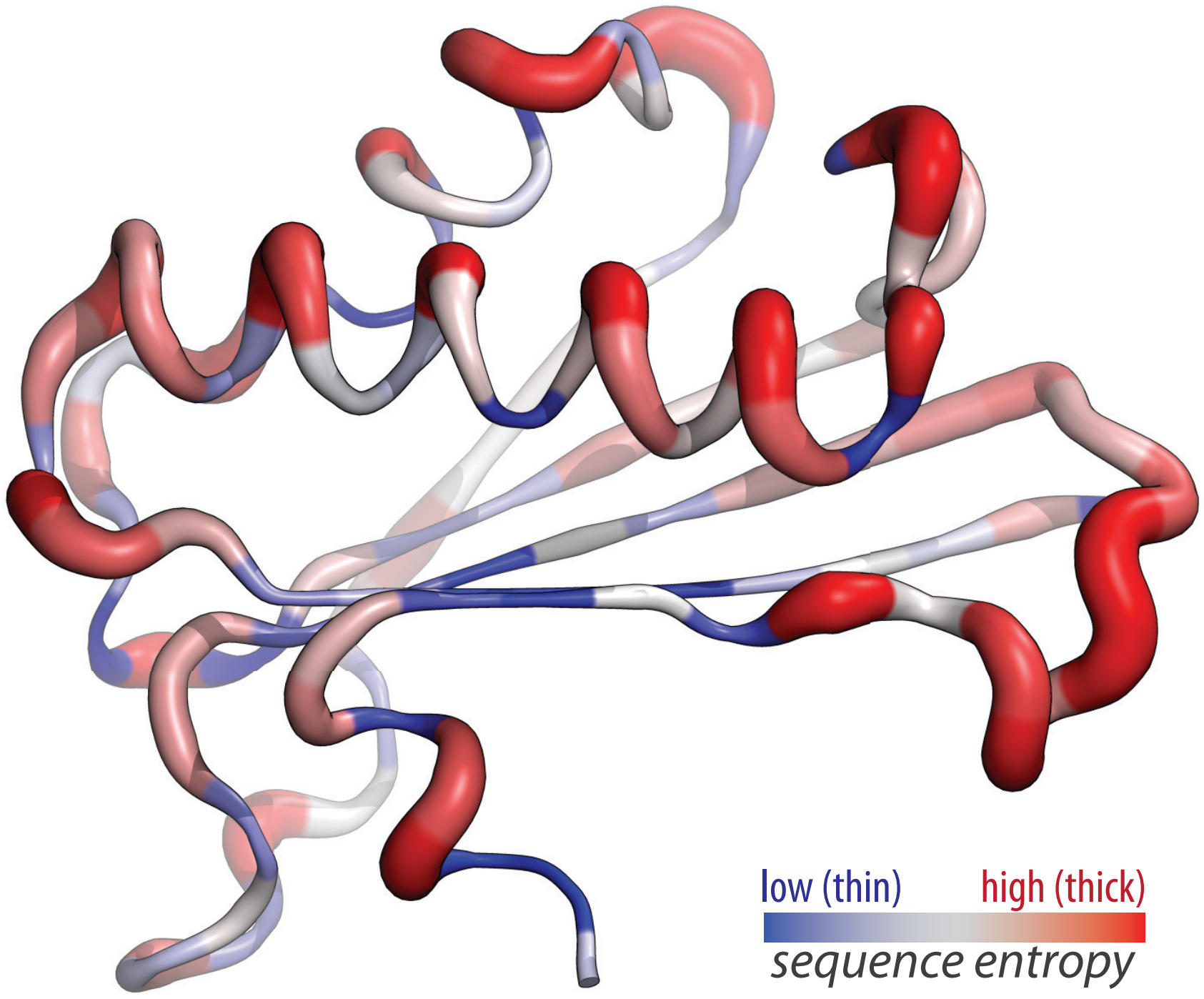

## 4. Summary

We have developed software for the processing and analysis of 3D structural data, utilizing abstractions and hierarchical data structures that are well-suited to representing macromolecular structures. The object-oriented code represents biopolymer structure *via* multiple classes and methods that together can be used to generate, process and store structural data; the software also provides efficient algorithms for manipulating coordinate-related data structures. Functionality for common structure analysis tasks is provided, including *(i)* file input/output (PDB parsers, writers); *(ii)* more advanced data structures, suitable for analysis of 3D coordinate sets (kD trees); *(iii)* structure analysis algorithms (surface areas, interatomic contacts); and *(iv)* a host of utility functions (structure clean-up, superposition, entity selection, data propagation). The software can also be interfaced to external libraries or used to 'wrap' binaries (*e.g.*, CCP4), thereby combining the efficiency of compiled languages like C or Fortran with the high level of abstraction and readability of an interpreted language such as Python. This latter aspect – using an interpreted language with an enhanced, interactive shell (Pérez & Granger, 2007) – facilitates exploratory data analysis and rapid prototyping of structural analysis workflows.

Originally motivated by needs arising in specific structural analysis projects (Cieślik & Derewenda, 2009), our software is not the only toolkit of this type (see, *e.g.*, *p3d* (Fufezan & Specht, 2009) or *Bio3d* (Grant *et al.*, 2006)). However, the code presented here is highly extensible, which is crucial for development of automated processing and structure analysis pipelines; in addition, relative to currently available packages, some aspects of the software (such as the kD tree implementation) provide rich functionality for neighbor search and related tasks (such as protein contact analysis). The emphasis on robust, fault-tolerant file processing and flexible data manipulation makes the code lightweight and concise, *versus* more ambitious Python packages that focus, for instance, on molecular simulations (*e.g.*, MMTK (Hinsen, 2000)). Perhaps the most useful feature of our software is not the code itself, but rather its context: In order to leverage the power of both sequence- and structure-based analysis, the software was fully integrated into the modern, feature-rich *PyCogent* bioinformatics framework. As illustrated above (§3.2), this symbiotic relationship enables sequence/structure analyses – *i.e.*, structural bioinformatics – to be pursued in a highly integrated, flexible and efficient manner.

**Acknowledgements** We gratefully acknowledge support from Univ of Virginia start-up funds (CM) and the Jeffress Memorial Trust (J-971; CM), as well as the NIH's NIGMS *via* the PSI2 Program (U54 GM074946-01; ZSD).

## Synopsis

Flexible abstractions, algorithms and data structures for protein structure analysis have been developed and integrated into the *PyCogent* bioinformatics software, thereby extending the scope of this sequence-based toolkit to enable combined sequence/structure-based bioinformatic studies.

Supplementary Material for

# Abstractions, algorithms and data structures for structural bioinformatics in *PyCogent*

Marcin Cieślik[1], Zygmunt S. Derewenda[2] & Cameron Mura[1]*

[1]Department of Chemistry, University of Virginia, Charlottesville, VA 22904-4319; [2]Department of Molecular Physiology & Biological Physics, University of Virginia Health Sciences Center, Charlottesville, VA 22908 USA
*cmura@virginia.edu

**Overview**

This supplementary material provides further details on our integration of 3D structure-handling capabilities into the Python-based *PyCogent* bioinformatics framework (Knight *et al.*, 2007) – in terms of basic software design decisions and conventions, as well as some of the routine aspects of using our new functionality for specific, structure-related tasks (PDB file manipulation, intermolecular contacts, *etc.*). As much as possible, each core principle described below is accompanied by brief, tangible 'use-cases', including examples of actual code (all in Python). The material is organized into several major sections, covering *(i)* The fundamental concept of '*entities*', their role in the software architecture, and their basic usage; *(ii)* A detailed example showing how to work with structures, particularly in terms of creating and manipulating entities; *(iii)* A brief case study, illustrating the analysis of interactomic contacts in a 3D structure; *(iv)* Two longer case studies, highlighting the utility of combining our structural tools with *PyCogent*'s existing feature set for sequence-based bioinformatics and phylogenetics; and *(v)* A final section pointing to further information. In essence, this document describes those portions of *PyCogent*'s application programming interface (API) of greatest relevance in performing basic or advanced structural bioinformatics analyses.

# Contents

# 1 Entities

The key concept of the '*Entity*' base class was introduced in the main text (§2). This section examines the central role of this class (and instances thereof – *i.e.*, '*objects*') in describing 3D molecular structure in *PyCogent*. We provide specific, concrete examples of working with *Entities*, in addition to samples of more advanced entity usage and manipulation.

## 1.1 What, *exactly*, is an 'entity'? – An object-oriented perspective

The '*Entity*' is the most basic class that provides methods specific to macromolecular structures. [Technically, each *Entity* is simply a Python dictionary (Python, 2010), with almost all built-in dictionary methods left untouched; *i.e.*, our *PyCogent Entities* inherit from Python's 'dict' class.] The '*Atom*', '*Residue*', '*Chain*', '*Model*' and '*Structure*' base classes all inherit from the '*Entity*' class. (In object-oriented parlance, they are said to be '*subclassed*'. As an analogy, *circles* are a subclass of *ellipses*, which comprise a subclass of the *conic sections* base class.) Despite their commnon derivation from the *Entity* class, there is some distinction between the behavior of these various classes. For instance, only an '*Atom*' entity cannot contain other entities – *e.g.* an instance of the '*Residue*' class can (and should) contain some '*Atom*' instances, but not *vice versa*. The methods common to container entities are within the '*MultiEntity*' class, from which all container entities inherit. The '*MultiEntity*' is also a subclass of the '*Entity*' base class. It is important not to directly create (or '*instantiate*') objects from the '*Entity*' and '*MultiEntity*' classes *de novo* (*versus* creating indirectly by, *e.g.*, using the built-in structure parser), because some crucial attributes (*e.g.*, position within the *SMCRA* hierarchy) are provided by the subclasses, and they are assumed to exist by any instance of the *Entity* class.

By parsing a macromolecular structure (*e.g.*, a PDB file) in *PyCogent*, one is implicitly instantiating a '*Structure*' entity and recursively populating it with a correct set of *atoms*, *residues*, and *chains* (and, possibly, multiple *models* from, *e.g.*, an NMR bundle).

## 1.2 Overview of methods and functions for selecting and grouping entities

The concept of *Entity* stems from that of structural hierarchy: *PyCogent* employs a hierarchical internal representation of macromolecular 3D structures, and 'entities' can be considered as the generic types of data structure elements at each level of the hierarchy (atoms, residues, chains, ...). A feature that distinguishes *PyCogent*'s approach to the handling of macromolecular structures is its flexible and concise way of defining entities, and then selecting, grouping and retrieving data from these objects. The remaining subsections (1.2.1 & 1.2.2) cover practical aspects of selecting entities *via* queries such as "*select all hydrogen atoms from chain B*", "*mask all hetero atoms*", "*remove all water molecules*", *etc*. We start with the high-level functions first (which are concise and standardized), and progress to lower-level methods for fine-grained manipulations to create entities that are comprised of precise collections of atoms. Indeed, the process of *entity construction* (*instantiation*) can be loosely thought of as the more familiar process of *atom selection* that is often performed in molecular modelling packages (see below).

### 1.2.1 An example: Selection based on structural hierarchy

This section illustrates the creation of entities based on groups of atoms. Most important, these "groups of atoms" are chosen not based upon the usual sorts of *atom selections* familiar to most users of molecular visualization software (*e.g.* "select myAtoms = chain A & resn Ser" in *PyMOL*); instead, the groups of atoms satisfy some particular, user-specified set of characteristics, in terms of the hierarchy of the data structures used to internally represent 3D molecular structures in *PyCogent*. Let's start by accessing a PDB file and creating a structure entity. We establish a connection to the PDB file server, download a file, and parse it:

```
>>> from cogent.parse.pdb import PDBParser
>>> from cogent.db.pdb import Pdb
>>> pdb = Pdb()                           # instantiates 'URLGetter' function to RCSB
>>> socket_handle = pdb['2E1F']           # ...and creates socket pointing to the 2E1F file
>>> structure = PDBParser(socket_handle) # ... and populate a new entity named 'structure'
```

Now, see what has been retrieved by querying various fields of the PDB header:

```
>>> print structure.header['name']
HYDROLASE
>>> print structure.header['experiment_type']
X-RAY DIFFRACTION
```

As expected (or not!), the structure is that of a hydrolase, determined from X-ray diffraction experiments. Now, how many chains does it have? (Note that there is only one 'model' in this file [it is not an NMR structure], hence the '`[(0,)]`' indexing in the following function call.)

```
>>> structure[(0,)].getChildren()    # N.B. 0-based indexing
[<Chain id=A>]
```

We found the 'A' chain of the first (0-based indexing) model. Yet, we can dig deeper:

```
>>> structure[(0,)][('A',)].sortedkeys()[0:2]
[(('H_HOH', 1, ' '),), (('H_HOH', 2, ' '),)]
```

Is the structure comprised only of waters ('`'H_HOH'`')? Probably not. Examine the contents of a chain by looking inside the dictionary of key/value pairs, retrieving the list of short ids (see below) and child entities:

```
>>> chain_A = structure[(0,)][('A',)]
>>> chain_A.keys()      # get the short_ids (output suppressed)
>>> chain_A.values()    # get the children  (output suppressed)
>>> len(chain_A)
147
```

Note, however, that this output value ('147') is an overestimate, as it counts water molecules in addition to amino acids. Continuing to re-type '`structure[(0,)][('A',)]`' is pretty boring and, additionally, it requires inspection of the number of models and chain IDs first (making it flexible, but less easily automatable). The *PyCogent* function that allows selection of atom-based entities from the hierarchy, based on their identity, is called '`einput`', and is illustrated below:

```
>>> from cogent.struct.selection import einput
>>> all_residues = einput(structure, 'R', 'my_residues')   # 'R' for residues
>>> all_atoms = einput(structure, 'A')                      # 'A' for atoms
>>> len(all_residues)
147
```

Still, waters are included. (Remember that they count as residues in PDB files!)

### 1.2.2 Extending the previous example: Property-based selection

We already have a collection of entities "all_residues" (see above) which contains all residues in the structure, regardless of the number of chains and models. Our task is to determine the number of non-water residues. The property that enables us to distinguish a water molecule from an amino acid is the residue name, which is stored for each residue as the "*name*" attribute. Therefore, try the following:

```
>>> chain_A.name
'A'
>>> first_child = chain_A.sortedvalues()[0]
>>> first_child.name
'H_HOH'
```

Now, a loop could be written to select those residues – looping over either the residues in ‘chain_A’, or the ‘all_residues’ set (they are the same):

```
>>> non_water = []
>>> for residue in chain_A:
...     if residue.name != 'H_HOH':
...         non_water.append(residue)
...
>>> len(non_water)
95
```

To make procedures such as the above more convenient, each entity (*e.g.*, a ‘*Chain*’ instance) has a method to select children based on a property. This method is known as ‘`selectChildren`’. Thus, a more compact equivalent of the above expression is:

```
>>> non_water = chain_A.selectChildren('H_HOH', 'ne', 'name').values()
```

. . . using Python’s builtin ‘`len`’ function to count the number of non-waters:

```
>>> non_water = all_residues.selectChildren('H_HOH', 'ne', 'name').values()
>>> len(non_water)
95
```

As illustrated in this example, the first argument of `selectChildren` is a value, and the second is an operator label (from Python’s ‘operator’ module; here ‘`ne`’ means ‘not equal’). The last argument (‘name’) is silently resolved by *PyCogent*’s ‘data_children’ method, which allows retrieval of data from a child entities attributes, *xtra dictionary*, or methods. In the above example, we get the data from the ‘name’ attribute. The `selectChildren` method returns a dictionary, where keys are the short ids and values are the child entities (these key/value pairs were alluded to in the above example too). The result can be put into a new entity holder:

```
>>> non_water_holder = einput(non_water, 'R')
```

However, having to first group the entities *via* the ‘`einput`’ function, then select them, only to have to put them into a new container (*i.e.*, the several steps described above) seems awkward. The same effect can be achieved more efficiently, in one step, using *PyCogent*’s ‘`select`’ function (from the struct.selection module) as follows:

```
>>> from cogent.struct.selection import select
>>> non_water_holder = select(structure, 'R', 'H_HOH', 'ne', 'name')
>>> len(non_water_holder)
95
```

Finally, we will use this functionality to ask if the structure contains a serine:

```
>>> serines = select(structure, 'R', 'SER', 'eq', 'name')
>>> serines.sortedkeys()[0]
('2E1F', 0, 'A', ('SER', 1146, ' '))
```

The function call would have raised a Python ‘ValueError’ if no entities could be selected.

### 1.2.3 Biochemical entities

The *SMCRA* hierarchy is currently aware of monomers (residues) and polymers (covalent chains of residues), as these are the generic structural units from which biomolecular structures (consisting of monomers) and assemblies (consisting of polymers) are constructed. Biophysical and biochemical data about atoms and standard amino acid residues are currently available in the code, and are planned for nucleotides. In terms of other, non-standard biochemical entities, only the most common ligands (from the virtually limitless set of possible small molecules observed in crystallographic structures) are currently ‘understood’ by the code;

other entities do not cause any trouble, they are simply treated as generic monomers or polymers ('generic' in the sense of our data-structures).

# 2 Working with molecular structure: Hierarchy, identity, entity

This section presents specific, highly detailed examples of how to work with macromolecular structures – *i.e.*, 3D coordinate files and coordinate-derived data. Much of the functionality described in this section traces its origins to an earlier *ZenPDB* module for Python (Cieślik and Derewenda, 2009). At the current stage of development, *PyCogent* input and output is limited to the Protein Data Bank (PDB; Bernstein *et al.* (1977)) file format, in addition to some close derivatives such as the 'PQR' format (Dolinsky *et al.* (2004); see also Mura (2010) for a usage note). Regardless of specific file format issues, note that, at the low level of software data structures and abstractions, our *PyCogent* functionality and 'internal' representations are entirely agnostic of file format, meaning that our structure-handling software is extensible, and other parsers could be easily added in the future.

## 2.1 Hierarchy & identity

Macromolecular structures are naturally described *via* hierarchical representations. Reflecting the underlying chemical organization and physical length scales spanned by biomolecules and their assemblies, the aforementioned atoms → residues → chains → models → structures → crystals hierarchy is composed of *entities* at each level. In such a representation, models are a subset of structures (*e.g.*, several NMR models of one protein structure), residues are seen as collections of atoms, and so on. Hierarchical representations of structure that are formulated in this manner are well-defined – *i.e.*, each entity is uniquely identified for a given structure, meaning that it has a unique *identifier*. We will refer to this unique identifier as the '**full_id**', in contrast to the '**short_id**' or just the '**id**' (which uniquely defines an entity only within the scope of its parent). Note that each entity has but a single parent, and may have multiple children (corresponding to multi-furcating branches in the tree-like hierarchy) – *e.g.*, a residue is necessarily part of only one peptide chain, but contains multiple atoms.
The following are examples of **full_id**'s:

```
# for an atom
('4TSV', 0, 'A', ('ARG', 131, ' '), ('O', ' '))
# for a residue
('4TSV', 0, 'A', ('ARG', 131, ' '))
# for a chain
('4TSV', 0, 'A')
```

Reading from right to left (*i.e.*, *S*←*M*←*C*←*R*←*A*), the first example identifies the oxygen ('*O*') atom from the peptide bond of the Arg131 residue, in the 'A' chain of the first model ('0') in the structure available from the PDB as '4TSV'. A short version of an **id** would look similar, but would be specific only at the level of its parent (*e.g.*, for an atom within a residue, or a chain id within an asymmetric unit [ASU]). Similarly, "('1T6Z', 0, 'B', ('PHE', 218, ' '), ('CE2', ' '))" would identify the 'CE2' atom of residue 'PHE218' in chain 'B' of the first model of structure '1T6Z'. In the example below, we see the **short_id** of the same atom. Of course, this information is not enough to uniquely pinpoint an atom in the structure, but does suffice to uniquely identify different atoms within the same residue.

```
(('O', ' '),)
```

As can be seen, the **full_id** is a linear tuple of **short_id**'s, which can be either a tuple (*e.g.*, '('O', ' ')' for an oxygen atom) or a string (*e.g.*, 'A' for chain A). All strings within a **short_id** have some special meaning – for instance, the id of a residue is structured as "('three letter AA name', residue_id, 'insertion code')".

It should be noted that PDB standards (http://www.wwpdb.org/docs.html) dictate that the 'residue_id' is that integer which should equal unity for the first residue in the naturally occuring protein. Thus, residues can have negative 'residue_id' values (consider, *e.g.*, the residues in an *N*-terminal affinity tag).

In terms of atomic IDs (full_id, short_id), we note that our software can handle sequence variability and structural microheterogeneity at the residue level – *i.e.*, our parsers and data structures recognize and adhere to PDB file-format conventions in terms of residue insertion codes (accounts for sequence numbering variability; see also §2.3.1 below) and alternate location ('altloc') indicators for multiple conformations of an individual residue. Both altlocs and insertion codes are preserved from the PDB file as part of the residue/atom IDs. As an example, the following block of code *(i)* fetches and parses a high-resolution structure (0.92 Å; 1L9L) that features several residues with alternate atom positions; *(ii)* instantiates model, chain, and residue objects from this struc; *(iii)* prints the occupancies and coordinates for the two altloc-related C$\beta$ atoms ('atom_A', 'atom_B') from residue R35:

```
>>> from cogent.parse.pdb import PDBParser
>>> from cogent.db.pdb import Pdb
>>> p=Pdb()
>>> struc = PDBParser(p['1L9L'])
>>> model = struc[(0,)]
>>> chain = model[('A',)]
>>> resi = chain[('ARG', 35, ' '),]
>>> atom_A = resi[("CB", 'A'),]       # conformer 'A' (major)
>>> atom_B = resi[("CB", 'B'),]       # conformer 'B' (minor)
>>> print 'A: occ = %0.2f; coords = %s' % (atom_A.occupancy, atom_A.coords)
A: occ = 0.64; coords = [ 13.036  11.48    4.875]
>>> print 'B: occ = %0.2f; coords = %s' % (atom_B.occupancy, atom_B.coords)
B: occ = 0.36; coords = [ 13.002  11.466   4.789]
```

The most detailed information about our software's dictionary-based approach to PDB file parsing – including proper handling of altlocs and insertion codes – can be found in the actual PDB parser ('`cogent/parse/pdb.py`' from the root of the *PyCogent* codebase).

## 2.2 Working with entities: A detailed treatment

Our first task will be to parse a PDB file and write a structure into an '*Entity*' hierarchy. This is not difficult and, for those already familiar with Python or *PyCogent*'s inner workings, should be quite straightforward. Of course, any PDB file can be used; the following example uses the '`4TSV.pdb`' file located in the `doc/data` directory of the *PyCogent* source code.
In a nutshell, the following is the easy (but implicit) way to achieve this – *Simply load the structure file*:

```
>>> import cogent
>>> structure = cogent.LoadStructure('data/4TSV.pdb')
```

The above two-liner involves quite a bit of magic. In the following, we acheieve the same result more manually and explicitly, to explore what data transformations are really occuring within the code. We begin by noting that the '`cogent.LoadStructure`' method really exists for the sake of convenience and efficiency, in order to allow one to get a structure object from a file in any of the supported input formats (currently, PDB [or closely related] format). Now, let's read and write the same PDB file by using the '`PDBParser`' and '`PDBWriter`' functions directly. The argument in the '`new_structure`' line can be any single '*Entity*' (*e.g.*, a '*Structure*' entity), or a container of entities (*e.g.*, a list of '*Atom*' or '*Residue*' entities):

```
>>> from cogent.parse.pdb import PDBParser
>>> from cogent.format.pdb import PDBWriter
>>> import tempfile, os
>>> pdb_file = open('data/4TSV.pdb')
```

```
>>> new_structure = PDBParser(pdb_file)
>>> open_handle, file_name = tempfile.mkstemp()
>>> os.close(open_handle)
>>> new_pdb_file = open(file_name,'wb')
>>> PDBWriter(new_pdb_file, new_structure)
>>> new_structure
<Structure id=4TSV>
```

In the above code listing, we first import the PDB parser and PDB writer, open a PDB file and parse the structure. You can verify that the 'PDBParser' does not close the open 'pdb_file':

```
>>> assert not pdb_file.closed
>>> assert not new_pdb_file.closed
```

At present, the 'PDBParser' automatically parses a lot information from the header of the PDB file and the atomic coordinate lines (it neglects anisotropic *B*-factors). Additional information is stored in the 'header' attribute of the returned object, which is a Python dictionary.

```
>>> structure.id      # the static id tuple.
('4TSV',)
>>> structure.getId() # the dynamic id tuple, use calls to get_id whenever possible.
('4TSV',)
>>> structure.getFull_id()     # only for a Structure-level entity is the 'full_id'
                               # equivalent to the 'id'
('4TSV',)
>>> structure.header.keys()    # the pdb header is parsed into a dictionary,
                               # as the header attribute
['bio_cmx', 'uc_mxs', 'name', 'solvent_content', 'expdta', 'bio_mxs',...
>>> structure.header['id']     # this is the 4-char PDB ID parsed from the header
                               # and used to construct the structure.id
'4TSV'
>>> structure.header['expdta'] # if this is 'X-RAY', we're probably dealing with an
                               # x-ray structre, and thus a lot crystallographic data
                               # is contained in the header.
'X-RAY'
```

Note that not *all* information from the PDB header is currently processed, due to intrinsic limitations of line-by-line text parsing; see 'cogent/parse/pdb.py' in the codebase to see exactly what fields are parsed. Thus, if you are interested in some particular data that is not parsed, you can access the unparsed ('raw') header *via* the 'raw_header' attribute (the same is true for the trailer); in doing so, note that the data will be chunked at the level of indidivual lines.

```
structure.raw_header
structure.raw_trailer
```

The *Structure* entity is a container for *Model* entities; the structure is just a dictionary of models:

```
>>> structure.items()
[((0,), <Model id=0>)]
>>> structure.values()
[<Model id=0>]
>>> structure.keys()
[(0,)]
>>> first_model = structure.values()[0]  # we name the first(and only)
                                         # model in the structure
>>> first_model_id = first_model.getId()
```

*PyCogent* provides more specific methods to work with entities. A useful method to access the contents of an entity is 'getChildren'. The optional argument to this method is a list of ids (*e.g.*, to access only a subset of children). More concise and powerful methods to work with entity children will be introduced as necessary.

```
>>> structure.getChildren() # the output should be the same as structure.values()
[<Model id=0>]
>>> children_list = structure.getChildren([first_model_id])
```

The '*MultiEntity*' generalizes the '*Entity*' concept (both of these may be viewed as base classes; see also §1.1). As an extension of the entity data structure, the *MultiEntity* provides methods shared between all subsidiary entities, and is technically a Python dictionary comprised of children entities (*i.e.*, entities that exist lower in the object hierarchy). A typical way to modify a particular property for all children in a *MultiEntity* would be to write a loop. In the following example, we change the name of every residue to 'UNK':

```
>>> some_model = structure.values()[0]
>>> some_chain = some_model.values()[0]
>>> for residue in some_chain.values():
...     residue.setName('UNK')
...
```

*PyCogent* allows these manipulations to be shortened. When a structure is created ('*instantiated*'), the top-level entity (*i.e.* the structure) gets a pointer list to all the entities it contains, the information being stored as a '`table`' attribute. For example, the structure entity will have a table with a list of all models, chains, residues, and atoms that it contains. The keys of this table are **full_ids** of the values of the actual entities. The table is divided into sections based on the hierarchy – *i.e.*, there are separate dictionaries for residues, atoms, chains, and models.

```
>>> sorted(structure.table.keys()) # all the different entity levels in the
                                   # table (which is a normal dictionary)
['A', 'C', 'M', 'R']
>>> structure.table['C']       # this is a full_id to entity mapping for
                               # all chains in the structure
{('4TSV', 0, ' '): <Chain id= >, ('4TSV', 0, 'A'): <Chain id=A>}
```

The creation of such a table is quite expensive, so it is created by default only for the structure entity. However, such tables can be generated as necessary (*e.g.,* for a chain, if needed):

```
>>> some_model = structure.values()[0]
>>> some_chain = some_model.values()[0]
>>> some_chain.setTable()
>>> # some_chain.table['R'] # all the residues
```

There is, however, a catch. Tables are not dynamic, meaning that they are not automagically updated whenever a child changes its id. This can be easily seen in the following example, where a new chain is created and a residue moved into it. A table is created for the chain, but it does not update the key after the child changes its name:

```
>>> from cogent.core.entity import Chain    # the chain entity
>>> new_chain = Chain('J')                  # instantiate an empty chain named 'J'
>>> new_chain.getId()
('J',)
>>> some_residue = structure.table['R'].values()[0]  # a semi-random residue from 'structure'
>>> # a possible output: <Residue UNK resseq=39 icode= >
>>> some_residue.setName('001')       # change the name to '001'
>>> # some_residue.getId()            # should return e.g. (('001', 39, ' '),)
>>> # some_residue.getFull_id()       # should return ('4TSV', 0, 'A', ('001', 39, ' '))
>>> new_chain.addChild(some_residue)  # move from chain 'A' in 4TSV into chain 'J'
>>> # new_chain.keys()                # should return: [(('001', 39, ' '),)]
>>> new_chain.setTable()
>>> # new_chain.table['R'].keys()     # should return: [('J', ('001', 39, ' '))]
>>> some_residue.setName('002')       # change the name to '002'
>>> # new_chain.keys()                # should return: [(('002', 39, ' '),)] # updated!
>>> # new_chain.table['R'].keys()     # should return [('J', ('001', 39, ' '))] not updated
>>> new_chain.setTable(force =True)   # update table
>>> # new_chain.table['R'].keys()     # should return [('J', ('002', 39, ' '))] updated
```

In handling these data structures, it is important to realize that Python dictionaries are unsorted, so the key order in two otherwise equal dictionaries will not be the same. Each time a child is changed in a way that affects the parent (*e.g.*, some part of the child's id is changed), the parent dictionary will be updated, and the order might also change. Therefore, one should **never** assume that an entity has a particular order.

```
>>> some_residue = some_chain.values()[0]
>>> old_id = some_residue.getId()      # e.g. (('ILE', 154, ' '),)
>>> some_residue.setName('VAL')
>>> new_id = some_residue.getId()      # e.g. (('VAL', 154, ' '),)
>>> some_chain.getChildren([old_id])   # nothing... not valid anymore
[]
>>> # some_chain.getChildren([new_id]) # e.g. [<Residue VAL resseq=154 icode= >]
```

But, an entity's *table* (which, remember, is accessed as an attribute of the entity object) is static, and does not get updated.

```
>>> some_full_id = some_residue.getFull_id()  # entities in tables are stored using their full ids!!
>>> # some_chain.table['R'][some_full_id]     # should raise a KeyError
>>> some_chain.setTable()                     # we make a new table
>>> some_chain.table['R'][some_full_id]
<Residue VAL resseq=131 icode= >
```

It is important to note that the *table* is a simple Python dictionary, and entity-specific methods like 'getChildren' are not available. You can figure out whether the table is up-to-date by examining its 'modified' attribute:

```
>>> some_chain.modified
False
```

If a result of 'True' had been returned, then some residue has been modified; in such instances, the 'setTable' or, in some cases, 'updateIds' methods should be applied as follows:

```
>>> some_chain.setTable()
>>> some_chain.updateIds()
```

Note that unnecessarily invoking those methods is not advised, as they may take some time.
The loop to run a child method can be (implicitly) circumvented by using the 'dispatch' method. This function calls (dispatches) the method for every child:

```
>>> some_model = structure.values()[0]
>>> some_chain = some_model.values()[1]
>>> some_chain.dispatch('setName', 'UNK')
>>> some_chain.modified
True
```

The above method has exactly the same effect as the loop – All residues within the chain will have their name set to 'UNK'. You can verify that the ids and dictionary keys were updated:

```
some_chain.keys()[0]   # output random e.g. (('UNK', 260, ' '),)
some_chain.values()[0] # e.g. <Residue UNK resseq=260 icode= >
```

## 2.3 More on entities

The following subsections provide additional information on three key concepts relating to our abstractions, data structures, and algorithmic implementations – namely, composite IDs (2.3.1), attribute dictionaries (2.3.2), and universal entity selection and input (2.3.3).

### 2.3.1 The composite id

Entities in *PyCogent* are uniquely identified by a composite id. Implemented as a tuple, the composite id is immutable, which makes it appropriate for use as a Python dictionary key. Each entity has two ids – the static 'short_id' and the on-demand, recursively-generated composite id. The form of the static id depends on the class inheriting from the *MultiEntity*, and is based on the PDB file standard; *e.g.*, the *Structure* class has a static id in the form ('name',), which is a tuple of unitary length, and with an arbitrary string as the structure name. By default the 'name' of a structure is its four-character PDB accession code. The static id of an instance of the *Residue* class is of the form (('residue_name', 'residue_id', 'insertion_code'),); see also the discussion of this topic in §2.1. The 'residue_name' is a three-char name of the residue (*e.g.*, 'ALA' for alanine, or '_DA' for deoxyadenosine in a nucleic acid); *PyCogent* is compatible with PDB v2.3 and v3.1 residue names. The 'residue_id' is the number of the residue in a chain, and, together with the insertion code (which reflects residue differences/insertions/deletions relative to the numbering in a reference [homologous] structure), should uniquely identify the position of the residue within a polypeptide. As alluded to above, 'short_id's identify entities unambiguously only at the level of their parents (atom ids are unique within residues, residue ids are unique within chains, *etc.*), not at the global level of a complete structure (*e.g.*, if a crystallographic ASU contains multiple chains of the same protein, the structure will have multiple residues with identical short ids). The unique composite id is generated from the id of an entity and all the ids of its parents in the hierarchy; thus, the 'full_id' of an atom derives from its 'short_id' and the 'short_id's of its parents at the residue, chain, model, and structure levels. The composite id is not a static attribute of an entity, because it depends also on its location within the hierarchy. It is combined on-demand from 'short_id's retrieved recursively from the parent. The decision to use a dictionary data structure, featuring the unique composite id as a key, provides an inherent safeguard against the possibility of generating ambiguous, poorly-structured PDB files or other output.

### 2.3.2 The attribute dictionary

The attribute dictionary is a useful design pattern, borrowed from the 'Bio.PDB' module of *Biopython* (Hamelryck and Manderick, 2003). It is based on the two ideas that *(i)* Each entity has a dictionary of additional, arbitrarily-specifiable properties (we termed this dict 'xtra'); and *(ii)* Instances of classes that assign properties to entities serve, themselves, as a mapping of their ids to the corresponding properties. While 'Bio.PDB' and *PyCogent* do not share code, both implementations take the similar approach of structure abstraction and representation *via* the *SMCRA* hierarchy. The '*xtra*' attribute provides a useful mechanism by which to store additional molecular properties (*e.g.*, a scalar-valued accessible surface area). Each property is structured as a name (in the form of a string) and some data value, which naturally corresponds to key/value pairs in the '*xtra*' dictionary. Notably, the properties can be an valid Python data structure. Upon instantiation, all assigning classes inherit from the *Entity* container object, which itself inherits from Python's built-in dictionary class; values are computed and assigned onto residues during instantiation. The finished instance is a mapping of the composite ids of entities, which have been modified, and their corresponding new property values. This results in both a unified API and larger flexibility for the end-user.

### 2.3.3 The universal entity selection and input

Structural analysis often requires one to select a particular part of the entities in a structure, perhaps to retrieve some particular property. For example, comparing amino acid compositions of a protein core and surface may involve partioning residues into two groups based on exposed surface area. Similarly, to compute overall protein charge, elementary residue charges must be looked-up and summed. *PyCogent* enables one to execute these sorts of calculations in a facile and uniform manner, *via* the inclusion of a special set of methods shared by all children of the *MultiEntity* class. The '`getData`' method retrieves, as a list, some arbitrary (user-specified) data from all children of a given entity. Additional wrapper methods, such as '`countChildren`' or '`freqChildren`', can operate on these data to provide histograms or frequencies of the observed values. Residues can also be selected based on their property values, and this can then be used to process atoms by removal (*e.g.*, hetero-atoms based on the 'h_flag' attriute, '`stripChildren`' method), splitting into groups (*e.g.* by residue type, '`splitChildren`' method) or ornament entities with their values (*e.g.*, contact area *via* the '`ornamentChildren`' method). The latter functionality can be used, for example, to sort residues by attribute values. (Known as the "Schwartzian transform", this *ornament* $\rightarrow$ *sort–by–ornament* $\rightarrow$ *de-ornament* idiom is common in Perl and Python.) The universal input function '`einput`' (defined in the 'struct/selection.py' module) converts any grouping of entities to an entity *holder* of a given level. The grouping can be either a list or an entity. The `einput` function will determine whether parent or child entities are requested. Each function or class which operates on entities will convert the input using `einput`, and internally uses the entity holder only; this renders the API more consistent and flexible. It is frequently necessary to perform a task on each atom or residue in a structure, *e.g.*, translate a subset of atomic coordinates by a given vector. This can be accomplished in *PyCogent* without having to resort to a nested series of loops (as is commonly done): An atom- or residue-level holder can be created *via* the `einput` function, followed by a call to the '`dispatch`' method of the holders (see also the `dispatch` method near the end of §2.2). Finally, the '`move`' method (or its special *MultiEntity* variant, '`moveRecursively`') can be applied, with the desired translation vector supplied as an argument.

# 3 Code performance; Platform independence; Data visualization

## 3.1 Code performance – Timing statistics

We have gauged the performance of our code, as integrated into *PyCogent*, by computing timings for two different types of operations: *(i)* parsing/loading a PDB file and *(ii)* calculating interatomic contacts in macromolecular structures. These timing statistics are shown as two plots (below): *(i)* Figure 1's file-loading time *vs* PDB file-size and *(ii)* Figure 2's interatomic calculation time *vs* PDB file-size. Notably, our software successfully parsed each PDB file in the random set of 2,886 non-redundant structures used to compute the timing data in Figures 1 & 2.

To assess the relationship between PDB file-size (structure size) and loading performance, we computed and plotted statistics for file parsing/loading time *versus* PDB file-size (measured as number of atoms, '$n$', in the structure). The data, shown in Figure 1, are consistent with the fact that structure files are parsed in linear-time (line-by-line); the slight nonlinearity for very large $n$ (note the linear-log scales) can be explained by the fact that atoms are validated by our code (*i.e.*, checked for name-clashes with existing atoms).

We have also assessed the influence of structure size upon contact calculation performance. Interatomic contact calculation in PyCogent is very versatile, and the performance of different queries will differ (*i.e.*, no universal gold standard to benchmark against). Thus, to give a rough estimate of the performance of our code implementation, we benchmarked the default setting – *i.e.*, a search for contacts between different chains in the asymmetric unit. The timings shown below (Figure 2) include the overhead required for initial construction of the kD-tree (which is $\mathcal{O}(n\log^2 n)$; note that subsequent range queries should be sub-linear.

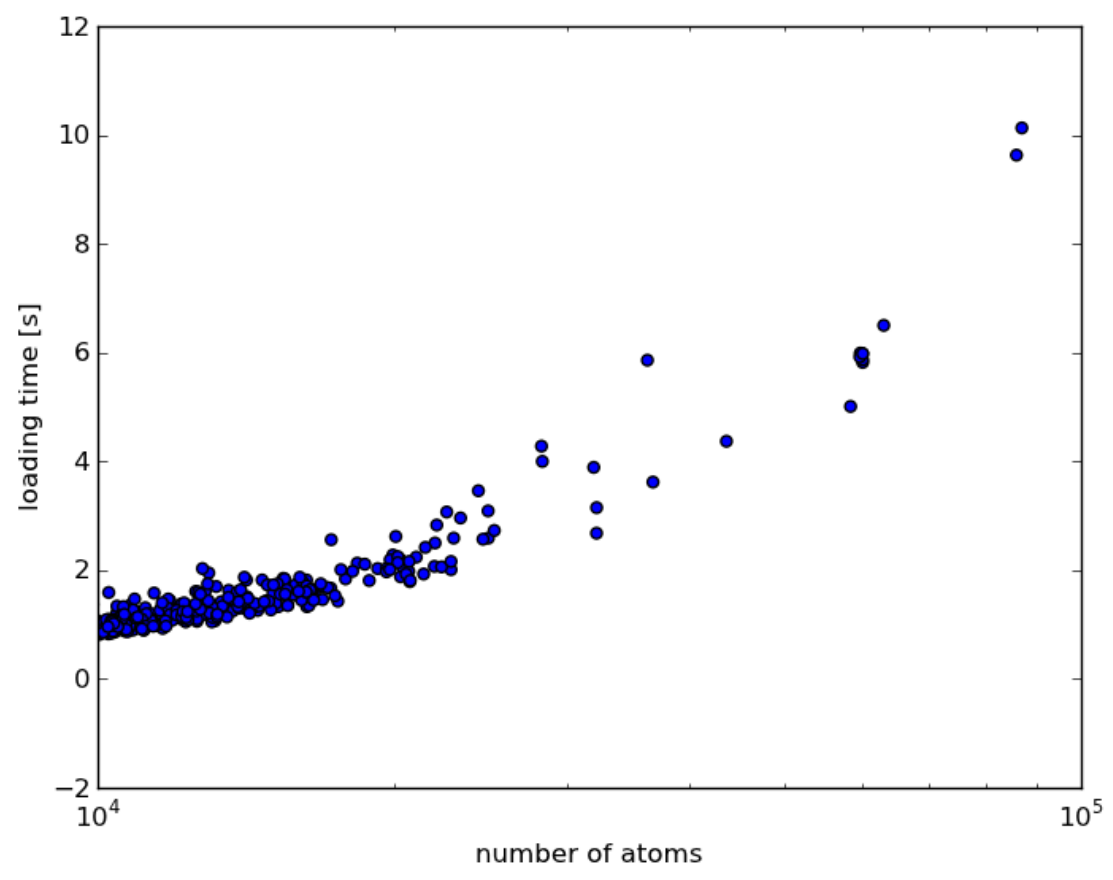

**Figure 1:** Measured loading times [s] *versus* number of atoms [‘$n$’] for a random set of 2,866 structures, ranging from 10,000 → 100,000 atoms.

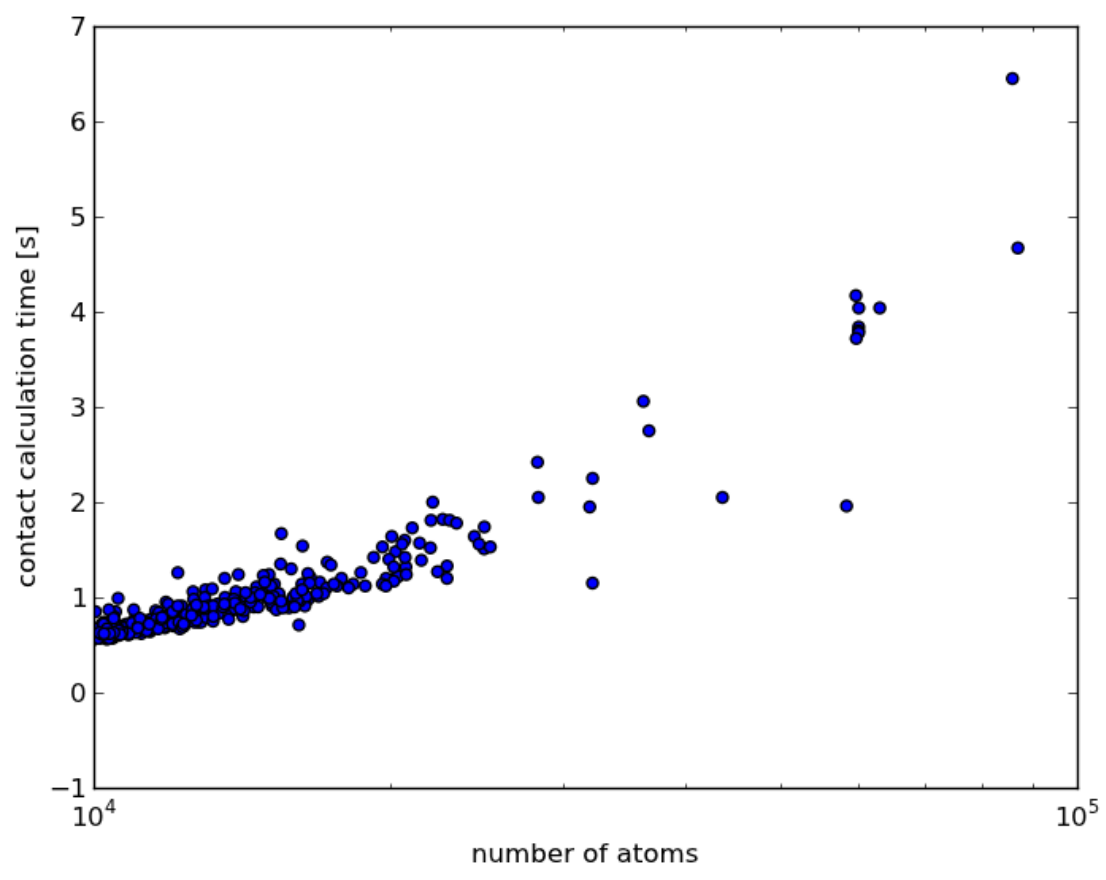

**Figure 2:** Contact calculation times [s] *versus* number of atoms [‘$n$’] for a random set of 2,866 structures, ranging from 10,000 → 100,000 atoms.

## 3.2 Notes on platform independence

The software introduced here was written in CPython, which is the standard/default, cross-platform implementation of Python. In much the same way as *PyMOL* or other Python-based software packages, our *PyCogent*-integrated code derives its platform-independence from the facts that *(i)* Python is platform-independent and *(ii)* apart from the Cython code (see paragraph below and main text), our software is written in clean Python (*i.e.*, no special tricks, customizations, add-on modules, or other dependencies were introduced beyond the defintion of the core language). Thus, the software can be used on any Python-capable platform (Linux/Unix, Mac, Windows); in principle the code should be cross-compatible with any platform featuring a fairly recent Python distribution (most modern platforms) and a C compiler (such as the free gcc) for the few Cython-based extensions.

As described in the main text (§2.3), numerically-intensive routines such as contact and surface area calculations were written as C extensions to Python, using the well-established, Pyrex-based Cython (.pyx files in the codebase). These files are translated into cross-platform .c source files that become “binary modules” upon compilation with the system-specific C compiler (*e.g.*, gcc), and that can then be directly imported in the Python interpreter (during installation this is all done automatically, not manually by the end-user); the compilation of .c files occurs at installation time. Also during installation (as during development and testing), the full software test-suite is run on all supported platforms, and if errors are spotted they are corrected. Though we have not encountered any reports of incompatibilities thus far, we would be happy to address any problems found by users on non-Linux platforms. It is important to emphasize that incompatibilities could result from bugs anywhere in the installation routine (*i.e.*, Cython sources → C sources → binary module), and only the first two elements of this tool-chain are within the scope of our *PyCogent*-based software.

## 3.3 Data visualization – Using our code with PyMOL

The ‘*Display the Quaternary Structure...*’ section of the main text (§3.1) illustrates the usage of our software in conjunction with *PyMOL*, for purposes of visualization. Because our code is integrated into the Python-based *PyCogent* package, and because the heavily Python-based *PyMOL* supplies a Python interface at the command prompt, such *PyCogent–PyMOL* integration is relatively seamless: One need simply ‘`import`’ the desired *PyCogent*-based Python modules into an active *PyMOL* session in order to access all of the functionality reported in our manuscript. This, indeed, is what is performed in the first line of the code-block in the main text ("from cogent.parse.pdb import PDBParser"). This process can be automated, making every *PyMOL* session *PyCogent*-aware, by including the above 'import' statement (or any similar statement) in one's *PyMOL* resource config file ('~/.pymolrc'); the contents of this file are read upon initialization of each new *PyMOL* session. Further information on this file (its location, usage, *etc.*) on the Linux/Unix, MacOS X and Windows operating systems may be found at http://www.pymolwiki.org/index-.php/Launching_PyMOL and http://www.pymolwiki.org/index.php/Pymolrc.

# 4 A brief case study: Interatomic contacts

## 4.1 Getting started with contacts

This section illustrates how to use various *PyCogent* modules (*e.g.*, struct, parse.pdb) and submodules (*e.g.*, contact, PDBParser) in order to identify inter-atomic (or, more generally, inter-*entity*) contacts between macromolecular entities such as individual amino acid residues, chains, *etc*. The method works within the context of a crystal or not, and presumes that the following boilerplate code is entered at the Python interpreter ('>>>'); the following code snippet includes the necessary 'import' statements, and creates a working set of *Entity* instances:

```
>>> from cogent.struct import contact
>>> from cogent.parse.pdb import PDBParser
>>> from cogent.struct import selection, manipulation, annotation
>>> pdb_fh = open('data/1HQF.pdb')       # read PDB file to open a filehandle
>>> pdb_structure = PDBParser(pdb_fh)    # ...and parse the structure file
>>> mymodel = pdb_structure[(0,)]        # create an object called 'model'
>>> chainA = mymodel[('A',)]             # ...and divide-up by chains
>>> chainB = mymodel[('B',)]
>>> chainC = mymodel[('C',)]
```

## 4.2 Exploring contacts within an asymmetric unit

In the following, we use the 'contact.contacts_xtra' function to identify interatomic contacts and save the results in the 'xtra' dictionary of an *Entity* instance. [Note: This function call is actually a wrapper around a lower-level function responsible for computing the contacts; the input data structure is annotated with the results of this function call, and a Python dictionary of these annotations is returned.] In the following invocation of this function, we do not search for contacts across the crystal lattice (*i.e.*, inter-unit cell contacts), but rather seek contacts between chains within a single asymmetric unit (ASU); searching for intra-ASU contacts is indeed the default behavior of the contact.contacts_xtra function, and this behavior can be altered *via* adjustable parameters (the argument 'contact_mode' may take values 'diff_cell' [inter-unit cell], 'diff_sym' [inter-ASU], or 'diff_chain' [inter-chain; the default]). In the particular example that follows, we search for all contacts between all atoms in the ASU, subject to the constraints that the atoms are at most 5.0 Å apart and that they belong to different chains:

```
>>> conts = contact.contacts_xtra(mymodel, xtra_key='CNT', search_limit=5.0)
```

How does one interpret the output of this function call? In what form are the resultant data structured? To explore these questions, note that the keys in the resulting 'conts' dictionary correspond to all those atoms that are involved in at least one (inter-chain) contact. The length of the dictionary is not the total number of contacts.

```
>>> atom_id = ('1HQF', 0, 'B', ('VAL', 182, ' '), ('C', ' '))
>>> atom_id_conts = conts[atom_id]
```

The value for this is a Python dictionary (see http://docs.python.org/tutorial/datastructures.html#dictionaries), where the contacts are stored in the given 'xtra_key' (the name of which defaults to 'CONTACTS', but which was set to 'CNT' in the step above).

```
>>> atom_id_conts
{'CNT': {('1HQF', 0, 'A', ('GLY', 310, ' '), ('CA', ' ')): (4.57341196482, 0, 0)}}
```

The value is a dictionary of contacts, with dictionary *keys* being ids of the involved atoms, and *values* being tuples consisting of *(i)* the distance in Å, *(ii)* symmetry operation id, and *(iii)* unit cell id. Note that, for contacts within an ASU, there are by definition no symmetry operations or unit cell translations relating the contacts, so both those values are returned as '0' (right-hand side of the above output line).

# 5 Two additional case studies

As an extension of the use-cases mentioned in the main text, this section illustrates the utility of combining our new structure-based functionality with *PyCogent*'s existing repertoire of sequence-based bioinformatic tools. The first use-case illustrates how sequence conservation information can be extracted from an alignment and mapped onto a structure, while the second example extends this by showing how coordinate-based data (in particular, 2° structural information) can be combined with the analysis of sequence entropy.

## 5.1 Use-case 1: Computing sequence entropy & mapping onto structure

### 5.1.1 Description of the use-case

This use-case illustrates how data computed from an alignment can be mapped onto a structure, *e.g.*, for purposes of visualization. The task is to identify and then illustrate the fragments of a protein structure that are conserved. Conservation will be measured in bits of entropy, which is a property that can be calculated from an alignment (see, *e.g.*, §11.2 of Durbin *et al.* (1998)).

The calculation ('`use-case_1.py`', below) proceeds as follows:

1. Calculate the Shannon entropy for each position in the alignment
    - Load an alignment in FASTA format.
    - Create a profile from the alignment.
    - Normalize the profile from amino acid counts to frequencies.
    - Compute the Shannon entropy from the normalized alignment.
2. Obtain protein sequences from the PDB entry
    - Load a structure in PDB file-format.
    - Create an ungapped sequence from the only chain in the structure.
    - Get the reference ID for the UniProtKB database.
    - Load the gapped UniProtKB sequence from the previous alignment.
3. Annotate the residues in the structure with bit-score values
    - Align the gapped and ungapped sequences, thereby mapping positions in the alignment to residues in the structure.
    - Prune those elements from the entropy array that correspond to gaps in the PDB-derived sequence,
    - Map the remaining values onto the structure's residues.

4. Save the data-annotated structure for visualization.

    - The bit-scores replace *B*-factors in the initial structure; many convenient visualization tools exist (*e.g.*, in PyMOL (DeLano, 2002)) for color-ramping a structure based on the *B*-factor field.

The non-trivial part of this task is to establish a valid mapping between the residues in the structure and positions in the alignment. Several issues contribute to this difficulty. First, the sequence for which the structure has been determined might not be identical to the sequence found in the database (*e.g.*, because of tags, mutated residues, *etc.*). Furthermore, not all residues in the crystallized sequence have defined coordinates; the ungapped chain sequence will simply miss those residues. To establish a proper mapping, a Needleman-Wunsch global alignment is computed between the ungapped sequence derived from the chain coordinates and the (almost identical, but gapped) sequence from the Pfam family alignment (see the code in the next section). Uncertainty bits for alignment positions, which correspond to gaps in the structure-derived sequence, are discarded because they cannot be visualized. The remaining bits are assigned to the corresponding residues in the structure. Finally, the structure is saved as a PDB file, with the *B*-factor column used to store the annotated results (*i.e.* the entropy bits).

### 5.1.2 Python code for this use-case

The following code for this use-case can also be found in the supplementary file ‘`use-case_1.py`’:

```
#!/usr/bin/env python
# import necessary modules from the Python standard library and
# PyCogent

from itertools import izip
import cogent
from cogent.core.annotation import SimpleVariable
from cogent.align.algorithm import nw_align
from cogent.struct.selection import select, einput
from cogent.struct.annotation import xtradata
from cogent.format.pdb import PDBXWriter

PDB_ID = '1UI9'
PFAM_ID = 'PF07736'

print "Executing use-case 1"
print "Loading alignment for family: %s" % PFAM_ID
alignment = cogent.LoadSeqs(PFAM_ID + '.fasta', moltype=cogent.PROTEIN, \
        label_to_name= lambda x:x.split('_')[0])

print "Creating Profile from alignment...",
profile = alignment.getPosFreqs()
profile.normalizePositions()
print "Valid: %s" % (profile.isValid(),)
print "Calculating Shannon Entropy..."
bits_gapped = [{'bits': b} for b in profile.rowUncertainty()]
print "Loading structure."
structure = cogent.LoadStructure(PDB_ID + '.pdb')
UPKB_ID = structure.header['dbref_acc']
chainA = einput(structure, 'C').values()[0]
print "Selecting chain A from structure."
upseq_gapped = alignment.getGappedSeq(UPKB_ID)
print "Selecting Sequence %s from alignment." % UPKB_ID
pdbseq_ungapped = chainA.getSeq()
print "Aligning sequences."
```

```
pdbseq_gapped, upseq_gapped = nw_align(pdbseq_ungapped, upseq_gapped)
print "Mapping residues to bits."
gaps = pdbseq_gapped.gapVector()
full_ids = pdbseq_ungapped.annotations[0].data
bits_ungapped = []
for idx, isgap in enumerate(gaps):
    if isgap:
        continue
    bits_ungapped.append(bits_gapped[idx])
full_id_bits = dict(zip(full_ids, bits_ungapped))
xtradata(full_id_bits, structure)
fn = PDB_ID + '_bfac_bits.pdb'
print "Writing result to new PDB file: %s" % fn
fh = open(fn, 'wb')
PDBXWriter(fh, structure, 'R', b_key='bits')
fh.close()
```

## 5.2 Use-case 2: Incorporating coordinate-derived data into the analysis

### 5.2.1 Description of the use-case

This use-case illustrates how results computed from coordinates can be used to further analyze the results of the previous example. The task is to test whether the entropy of a residue varies significantly depending on its secondary structure.
The steps are as follows:

1. Load the resulting structure from the previous use-case.
    - Annotate residues with data from the *B*-factor column of their constituent atoms.
2. Group bit-scores based on the residue secondary structure class.
    - Determine the secondary structure of residues in the structure using *Stride*.
    - Disregard residues for which no secondary structure information is present.
    - Group residues based on secondary structure class.
    - Obtain a vector of bit-values for each group.
3. Perform an analysis of variance (ANOVA) test to see if the between-group variances differ.

In the previous use-case, the results were saved as a PDB file in which the *B*-factor column was co-opted to hold bit-values (strictly speaking, each atom was assigned a *B*-factor value equal to the uncertainty of its parent residue, from the alignment-based profile). After parsing, this column became a per-atom attribute, and residues have to be back-annotated. The popular command-line application '*Stride*' is used to define the secondary structure of the residues in the structure, and residues are divided into groups based on the classes defined by *Stride*. If a residue has no annotation (*e.g.* because of a missing atom), it is omitted. For each of the groups, a vector containing the bit-values of the residues is created. These results are then used to perform a one-way ANOVA statistical test for differences.

### 5.2.2 Python code for this use-case

The following code for this use-case can also be found in the supplementary file '`use-case_2.py`':

```
#!/usr/bin/env python

import cogent
from numpy import mean, std
from cogent.app.stride import stride_xtra
from cogent.struct.selection import select, einput
from cogent.maths.stats.util import Numbers
from cogent.maths.stats.test import ANOVA_one_way

PDB_FILE = '1UI9_bfac_bits.pdb'

print "Loading structure: %s" % (PDB_FILE,)
structure = cogent.LoadStructure(PDB_FILE)
print "Running Strid"
print stride_xtra(structure)
print "Remove residues which have no secondary-structure annotation"
ss_residues = select(structure, 'R', None, 'ne', 'STRIDE_SS', xtra=True)
print "Transfer bfactor values from atoms to residues."
ss_residues.propagateData(mean, 'A', 'bfactor')
print "Group residues based on secondary-structure"
classes = ss_residues.splitChildren('STRIDE_SS', xtra=True)
print "Got secondary-structure groups: %s" % (" ".join(classes.keys()))
results = []
print "Get vectors of bits for the residues in each group."
for cls, ents in classes.iteritems():
    residues = einput(ents.values(), 'R')
    residue_bits = Numbers(residues.getData('bfactor'))
    results.append(residue_bits)
    print cls, residue_bits.Mean, '+/-', residue_bits.StandardDeviation

ar = ANOVA_one_way(results)
print "Result of the ANOVA one way test: F=%s , right tail=%s" % (ar[3], ar[-1])
```

# 6 For further information...

This document is intended to contain sufficient information so that the interested reader can *(i)* begin performing structural analyses in *PyCogent* and *(ii)* understand the inner-workings (and, therefore, capabilities and limitations) of *PyCogent*'s structure analysis tools. While we have attempted to be as thorough as possible, the treatment is not comprehensive. Also, note that much of the information found here also occurs in various places in the official, freely-available *PyCogent* documentation, located both online and as reStructuredText files in the `doc/{cookbook,examples}/` directories of the source-code. In addition, the following two files (both in the '`doc/cookbook`' subdirectory of the codebase) provide useful collections of information pertaining to structure analysis in *PyCogent*:

- `structural_data.rst` — A basic introduction to *PyCogent*'s structure-related functionality, that document is largely structured as a FAQ, addressing such questions as "*How do I retrive a structure from the PDB?*", "*How do I get a list of all residues in a chain?*", "*How do I calculate the distance between two atoms?*", *etc*; Python one-liners provide simple solutions to many such questions, so that guide offers a useful complement to the present document.

- `structural_data_2.rst` — Provides an introduction to more advanced structure-related work in *PyCogent*, involving many of the concepts and usage principles that have been introduced in the present document (*e.g.*, the *MultiEntity*, methods for selecting children, *AtomHolder* instances, and so on).